\documentclass[aps,prd,preprintnumbers,letterpaper,superscriptaddress,nofootinbib,floatfix]{revtex4-2}
\pdfoutput=1

\usepackage{graphicx}
\usepackage{microtype}
\usepackage{amsmath}
\usepackage{amssymb}
\usepackage{xcolor}
\usepackage{hyperref}
\usepackage{natbib}

\graphicspath{{Images/}}
\hypersetup{
    colorlinks=true,
    linkcolor=blue,
    citecolor=magenta,
    urlcolor=red
}

\begin{document}

\title{The Analytical Solutions of Dyonic Black Holes in Einstein-Euler-Heisenberg Theory}

\author{Haoyuan \surname{Luo}}
\email{2770521341@qq.com}
\affiliation{School of Science, Jiangsu University of Science and Technology, 212100, Zhenjiang, China.}

\author{Nan \surname{Cao}}
\email{1913562551@qq.com}
\affiliation{School of Science, Jiangsu University of Science and Technology, 212100, Zhenjiang, China.}

\author{Xiao Yan \surname{Chew}}
\email{xiao.yan.chew@just.edu.cn}
\affiliation{School of Science, Jiangsu University of Science and Technology, 212100, Zhenjiang, China.}

\author{Kok-Geng \surname{Lim}}
\email{K.G.Lim@soton.ac.uk}
\affiliation{University of Southampton Malaysia, 79100 Iskandar Puteri, Malaysia.}

\author{Chao \surname{Chen}}
\email{cchao012@just.edu.cn}
\affiliation{School of Science, Jiangsu University of Science and Technology, 212100, Zhenjiang, China.}

\author{Dong-han \surname{Yeom}}
\email{innocent.yeom@gmail.com}
\affiliation{Department of Physics Education, Pusan National University, Busan 46241, Republic of Korea}
\affiliation{Research Center for Dielectric and Advanced Matter Physics, Pusan National University, Busan 46241, Republic of Korea}
\affiliation{Leung Center for Cosmology and Particle Astrophysics, National Taiwan University, Taipei 10617, Taiwan }

\begin{abstract}
Recently, we constructed analytical purely electric and purely magnetic black-hole solutions in Einstein--Euler--Heisenberg theory, while the corresponding dyonic solutions were obtained numerically \cite{Luo:2026srx}. Motivated by the recent analytical construction of a dyonic black hole at the special coupling locus $b=a/2$ \cite{Ahmed:2026ufj}, we revisit the general dyonic sector. We show that exact analytical dyonic black-hole solutions can be constructed for nonlinear couplings satisfying $2b>a$, without imposing the restriction $b=a/2$. The mass function is expressed in closed form in terms of the Lauricella hypergeometric function. In particular, our construction yields an exact dyonic solution for the Euler--Heisenberg coupling $b=7a/4$. The solution of Ref.~\cite{Ahmed:2026ufj} is recovered in the limiting case $2b\rightarrow a$.
\end{abstract}

\maketitle


Recently, we obtained numerical solutions describing dyonic black holes
\cite{Luo:2026srx}. At the time of submitting our work to arXiv, we were not aware that it is possible to obtain analytical dyonic solutions.
Subsequently, inspired by Ref.~\cite{Ahmed:2026ufj}, we found that the dyonic
black holes can, in fact, be constructed analytically.

In Ref.~\cite{Ahmed:2026ufj}, a generic quadratic nonlinear electrodynamics
(NED) model coupled to Einstein gravity is described by the action
\begin{equation}
S=\frac{1}{16\pi}\int d^4x\,\sqrt{-g}
\left[R-\mathcal{L} (\mathcal{F},\mathcal{G})\right]\,,
\end{equation}
where the Lagrangian $\mathcal{L}(\mathcal{F},\mathcal{G})$ is given by
\begin{equation}
\mathcal{L}(\mathcal{F},\mathcal{G})
=
\mathcal{F}-
\left(
a\mathcal{F}^2+b\mathcal{G}^2
\right)\,.
\end{equation}
Here, $a$ and $b$ are the nonlinear coupling parameters, and
\begin{equation}
\mathcal{F}\equiv F_{\mu\nu}F^{\mu\nu}\,,
\qquad
\mathcal{G}\equiv \tilde{F}^{\mu\nu}F_{\mu\nu} \,,
\end{equation}
are the two electromagnetic invariants. The electromagnetic field tensor and
its Hodge dual are defined by
\begin{equation}
F_{\mu\nu}=\partial_\mu A_\nu-\partial_\nu A_\mu\,,
\qquad
\tilde{F}^{\mu\nu}
=
\frac{1}{2}\eta^{\mu\nu\rho\sigma}F_{\rho\sigma}\,,
\end{equation}
where $\eta^{\mu\nu\rho\sigma}$ is the completely antisymmetric Levi-Civita tensor. The equations of motion are given by
\begin{align}
R_{\mu\nu}-\frac{1}{2}g_{\mu\nu}R
&=
2\left(
\mathcal{L}_{\mathcal{F}}
F_{\mu\lambda}F_{\nu}{}^{\lambda}
+
\mathcal{L}_{\mathcal{G}}
F_{\mu\lambda}\tilde{F}_{\nu}{}^{\lambda}
\right)
-\frac{1}{2}g_{\mu\nu}\mathcal{L} \,,
\\
\nabla_\mu\left(
\mathcal{L}_{\mathcal{F}}F^{\mu\nu}
+
\mathcal{L}_{\mathcal{G}}\tilde{F}^{\mu\nu}
\right)
&=0\,,
\label{eq:EOM_gauge}
\end{align}
where
\begin{equation}
\mathcal{L}_{\mathcal{F}}
\equiv
\frac{\partial\mathcal{L}}{\partial\mathcal{F}}\,,
\qquad
\mathcal{L}_{\mathcal{G}}
\equiv
\frac{\partial\mathcal{L}}{\partial\mathcal{G}}\,.
\end{equation}

The following Ansatz employed to construct static and spherically symmetric black-hole solutions,
\begin{equation}
\label{eq:metric}
ds^2
=
-e^{-2\sigma(r)}N(r)dt^2
+\frac{dr^2}{N(r)}
+r^2\left(d\theta^2+\sin^2\theta\,d\varphi^2\right)\,,
\end{equation}
where, as in Eq.~(16) of Ref.~\cite{Luo:2026srx}, we set
\begin{equation}
\sigma(r)=0\,,
\end{equation}
and define
\begin{equation}
\label{eq:func_N}
N(r)=1-\frac{2m(r)}{r}\,. 
\end{equation}
The function $m(r)$ is the Misner--Sharp mass function, and the ADM mass $(M)$ is
obtained from its asymptotic value,
\begin{equation}
M=\lim_{r\rightarrow\infty}m(r)\,.
\end{equation}

The Ansatz for the static and spherically symmetric gauge potential is given by
\begin{equation}
A_\mu dx^\mu
=
V(r)\,dt+P\cos\theta\,d\varphi\,,
\end{equation}
where $P$ denotes the magnetic charge. The substitution of gauge field into Eq.~\eqref{eq:EOM_gauge} gives rise to
\begin{equation}
    \partial_r \left[ 4 a r^2 V'^3 + r^2 \left( 1+\frac{4cP^2}{r^4}  \right)   V' \right] = 0\,,
\end{equation}
where $c=2b-a$. The direct integration yields
\begin{equation}
    4a r^2 V'^3 + r^2\left(1+\frac{4cP^2}{r^4} \right) V' = - Q\,, \label{eq:E_cubic1}
\end{equation}
where $Q$ is the electric charge. We define the electric field $E(r) = - V'(r)$ to cast the above equation into the cubic equation,
\begin{equation}
E(r)^3+p_1(r)E(r)+p_2(r)=0\,,
\label{eq:E_cubic2}
\end{equation}
where
\begin{equation}
p_1(r)
=
\frac{1}{4a}
\left(
1+\frac{4 c P^2}{r^4}
\right)\,,
\qquad
p_2(r)
=
-\frac{Q}{4a r^2}\,.
\end{equation}
It possesses three roots, which are given by
\begin{align}
E_1(r)
&=
E_1^+(r)+E_1^-(r)\,,
\label{eq:root1}
\\
E_2(r)
&=
-\frac{1}{2}E_1(r)
+\frac{\sqrt{3}}{2}i
\left[E_1^+(r)-E_1^-(r)\right]\,,
\label{eq:root2}
\\
E_3(r)
&=
-\frac{1}{2}E_1(r)
-\frac{\sqrt{3}}{2}i
\left[E_1^+(r)-E_1^-(r)\right]\,,
\label{eq:root3}
\end{align}
where
\begin{equation}
E_1^\pm(r)
=
\sqrt[3]{
-\frac{p_2}{2}
\pm
\sqrt{
\left(\frac{p_2}{2}\right)^2
+
\left(\frac{p_1}{3}\right)^3
}
}\,.
\label{eq:discri}
\end{equation}
The discriminant $\Delta(r)$ of Eq.~\eqref{eq:E_cubic2} is defined as
\begin{equation}
\Delta(r)
=
\left(\frac{p_2}{2}\right)^2
+
\left(\frac{p_1}{3}\right)^3\,.
\end{equation}
Here we find that $\Delta > 0$ for $c > 0$. Therefore, $E_1(r)$ is the only physical solution. 


In Ref.~\cite{Luo:2026srx}, we derived a first-order ordinary differential equation (ODE) for the mass function $m(r)$. For the present configuration, Eq.~(19) of Ref.~\cite{Luo:2026srx} can be derived as
\begin{equation}
    m'(r) = \frac{P^2}{2 r^2}\left(1-\frac{2 a P^{ 2}}{r^4}\right) + \frac{r^2 E}{2} \left[\left(1 + \frac{4 cP^2}{r^4} \right) E + 6a E^3\right]\,.
\end{equation}
Using Eq.~\eqref{eq:E_cubic1}, this ODE can be simplified to
\begin{equation}
    m'(r) = \frac{P^2}{2 r^2} - \frac{aP^4}{r^6} + \left( \frac{Q}{2} + ar^2E^3 \right)E\,.\label{eq:ODE_m1}
\end{equation}

To integrate Eq.~\eqref{eq:ODE_m1} analytically, we introduce the auxiliary variable $z$ and the function $H(z)$,
\begin{equation}
    z \equiv r^2 E\,,\qquad H(z) \equiv az^2 +cP^2\,.
\end{equation}
In the Reissner-Nordstr\"om limit, $E(r)=Q/r^2$, and hence $z=Q$. In terms of $z$, the radial coordinate and the electric field can be expressed as
\begin{equation}
    r(z)^4 = \frac{4zH(z)}{Q-z}\,,\qquad E(z)^2=\frac{z(Q-z)}{4H(z)}\,.\label{eq:r_E_of_z}
\end{equation}
For $Q>0$, $c>0$, and the physical branch with $E(r)>0$, the nonlinear cubic term weakens the electric field relative to its RN value. Consequently, $0<z<Q$. Taking the positive real branches, we obtain
\begin{equation}
    r(z) = \sqrt[4]{\frac{4zH(z)}{Q-z}}\,,\qquad E(z)=\frac{1}{2}\sqrt{\frac{z(Q-z)}{H(z)}}\,.
\end{equation}
Meanwhile, substituting $E(r)=z/r^2$ into Eq.~\eqref{eq:ODE_m1} yields the more compact expression
\begin{equation}
    m'(r) = \frac{P^2+Qz}{2 r^2} + \frac{a\left(z^4-P^4\right)}{r^6}\,.
\end{equation}
Differentiating $r(z)$ with respect to $z$, we find
\begin{equation}
    \frac{dr}{dz} = \frac{r}{4} \frac{d}{dz}\ln\left[\frac{zH(z)}{Q-z}\right] = \frac{r}{4} \left[\frac{1}{z} + \frac{2az}{H(z)} + \frac{1}{Q-z}\right]\,.
\end{equation}
Combining this relation with Eq.~\eqref{eq:r_E_of_z}, the mass equation becomes
\begin{equation}
    \frac{dm}{dz} = \frac{dm}{dr}\frac{dr}{dz} = \frac{r}{4} m'(r) \left[\frac{1}{z} + \frac{2az}{H(z)} + \frac{1}{Q-z}\right] = \frac{\mathcal{P}_8(z)}{16\sqrt{2}z^{9/4}H(z)^{9/4}(Q-z)^{3/4}}\,.\label{eq:ODE_m2}
\end{equation}
Here, the numerator $\mathcal{P}_8(z)$ is an eighth-order polynomial,
\begin{equation}
    \mathcal{P}_8(z) = \sum_{n=0}^8 c_n z^n\,,\label{eq:poly_P_z}
\end{equation}
whose coefficients are listed in Table~\ref{tab:coeff}.
\begin{table}[htbp]
    \centering
    \begin{tabular}{c c c c}
        \hline\hline
        $n$
        & \makebox[0.2\linewidth][c]{$0$}
        & \makebox[0.2\linewidth][c]{$1$}
        & \makebox[0.2\linewidth][c]{$2$}\\
        \hline
        $c_n$
        & \makebox[0.2\linewidth][c]{$-acP^6Q^2$}
        & \makebox[0.2\linewidth][c]{$\left(ac+2c^2\right)P^6Q$}
        & \makebox[0.2\linewidth][c]{$\left(2c^2-3a^2\right)P^4Q^2$}\\
        \hline\hline

        $n$
        & \makebox[0.2\linewidth][c]{$3$}
        & \makebox[0.2\linewidth][c]{$4$}
        & \makebox[0.2\linewidth][c]{$5$}\\
        \hline
        $c_n$
        & \makebox[0.2\linewidth][c]{$\left(5a^2+8ac\right)P^4Q$}
        & \makebox[0.2\linewidth][c]{$-\left(2a^2 + 4ac\right)P^4+9acP^2Q^2$}
        & \makebox[0.2\linewidth][c]{$\left(6a^2-5ac\right)P^2Q$}\\
        \hline\hline

        $n$
        & \makebox[0.2\linewidth][c]{$6$}
        & \makebox[0.2\linewidth][c]{$7$}
        & \makebox[0.2\linewidth][c]{$8$}\\
        \hline
        $c_n$
        & \makebox[0.2\linewidth][c]{$9a^2Q^2-4a^2P^2$}
        & \makebox[0.2\linewidth][c]{$-9a^2Q$}
        & \makebox[0.2\linewidth][c]{$2a^2$}\\
        \hline\hline
    \end{tabular}
    \caption{Coefficients $c_n$ of the polynomial $\mathcal{P}_8(z)$ defined in Eq.~\eqref{eq:poly_P_z}.}
    \label{tab:coeff}
\end{table}

To express the antiderivative of Eq.~\eqref{eq:ODE_m2} in terms of a Lauricella function, we introduce
\begin{equation}
    \eta^2 \equiv -\frac{cP^2}{a} \,.
\end{equation}
The factors $H(z)$ and $Q-z$ can be rewritten as
\begin{equation}
    H(z)=cP^2\left(1-\frac{z}{\eta}\right)\left(1+\frac{z}{\eta}\right)\,,\qquad Q-z=Q\left(1-\frac{z}{Q}\right)\,.
\end{equation}
Notice that $\eta$ is generally complex when $ac>0$. Nevertheless, the two factors containing $\eta$ occur as a conjugate pair, and their product reconstructs the real function $H(z)$.

Substituting these factorizations and Eq.~\eqref{eq:poly_P_z} into Eq.~\eqref{eq:ODE_m2}, we obtain 
\begin{equation}
    \frac{dm}{dz} = B\sum_{n=0}^8 c_n  z^{n-9/4} \left(1-\frac{z}{Q}\right)^{-3/4} \left(1-\frac{z}{\eta}\right)^{-9/4} \left(1+\frac{z}{\eta}\right)^{-9/4}\,,\label{eq:ODE_m3}
\end{equation}
where 
\begin{equation}
    B=\frac{1}{16\sqrt{2}\left(cP^2\right)^{9/4}Q^{3/4}}\,.
\end{equation}


The Lauricella hypergeometric function $F_D^{(K)}$ admits the Euler-type integral representation
\cite{Kraniotis:2010gx, Lai:2020mfi, Duhr:2023bku} 
\begin{equation}
    F^{(K)}_{D} \left(\alpha;\beta_1,...,\beta_K;\gamma;x_1,...,x_K\right) = \frac{\Gamma(\gamma)}{\Gamma(\alpha)\Gamma(\gamma-\alpha)} \int_0^1 dt \, t^{\alpha-1} (1-t)^{\gamma-\alpha-1} \prod \limits_{i=1}^K \left(1-x_i t\right)^{-\beta_i}\,.
\end{equation}
For $K=3$ and $\gamma=\alpha+1$, this expression reduces to
\begin{equation}
    F^{(3)}_D \left(\alpha;\beta_1,\beta_2,\beta_3;\alpha+1;x_1,x_2,x_3\right)= \alpha \int_0^1 dt \, t^{\alpha-1}  \prod \limits_{i=1}^3 (1-x_it)^{-\beta_i}\,.
\end{equation}

To derive the required antiderivative, we perform the change of variables
\begin{equation}
    u \equiv zt\,,\qquad t=\frac{u}{z}\,, \qquad dt=\frac{du}{z}\,.
\end{equation}
The two integration limits: $t=0$ and $t=1$ correspond to $u=0$ and $u=z$, respectively. It follows that
\begin{equation}
    F^{(3)}_D \left(\alpha;\beta_1,\beta_2,\beta_3;\alpha+1;x_1 z,x_2 z,x_3 z\right)= \alpha \int_0^z \frac{du}{z} \, \left(\frac{u}{z}\right)^{\alpha-1}  \prod \limits_{i=1}^3 (1-x_i u)^{-\beta_i}\,.
\end{equation}
Consequently,
\begin{equation}
    \int_0^z du \, u^{\alpha-1}  \prod \limits_{i=1}^3 (1-x_i u)^{-\beta_i}=\frac{z^{\alpha}}{\alpha}F^{(3)}_D \left(\alpha;\beta_1,\beta_2,\beta_3;\alpha+1;x_1 z,x_2 z,x_3 z\right)\,.
\end{equation}
This identity has also been employed in Refs.~\cite{Akerblom:2004cg,Fathi:2023qyl}. Defining
\begin{equation}
    G(z) = \frac{z^{\alpha}}{\alpha}F^{(3)}_D \left(\alpha;\beta_1,\beta_2,\beta_3;\alpha+1;x_1 z,x_2 z,x_3 z\right) \,,
\end{equation}
the fundamental theorem of calculus gives
\begin{align}    
    G'(z) &= z^{\alpha-1}  \prod \limits_{i=1}^3 (1-x_i z)^{-\beta_i}\,,\label{eq:dG_dz}\\
    \int dz\, z^{\alpha-1}  \prod \limits_{i=1}^3 (1-x_i z)^{-\beta_i} &= \frac{z^{\alpha}}{\alpha}F^{(3)}_D \left(\alpha;\beta_1,\beta_2,\beta_3;\alpha+1;x_1 z,x_2 z,x_3 z\right)+C\,,
\end{align}
where $C$ is an integration constant and Eq.~\eqref{eq:dG_dz} can be also found in Ref.~\cite{Kaneko:2010gta}. The Euler integral provides a direct derivation when its convergence conditions are satisfied. Outside that domain, including the terms with $n=0$ and $n=1$ below, the antiderivative is understood through analytic continuation of the Lauricella function.


Comparing this identity with Eq.~\eqref{eq:ODE_m3}, we identify
\begin{equation}
    \alpha=n-\frac{5}{4}\,,\qquad \beta_1=\frac{3}{4}\,,\qquad \beta_2=\beta_3=\frac{9}{4}\,,\qquad x_1=\frac{1}{Q}\,, \qquad x_2=-x_3= \frac{1}{\eta}\,.
\end{equation}
Thus, the mass function can be written as
\begin{align}
    m(z) &= B\int dz \, \sum_{n=0}^8 c_n  z^{n-9/4} \left(1-\frac{z}{Q}\right)^{-3/4} \left(1-\frac{z}{\eta}\right)^{-9/4} \left(1+\frac{z}{\eta}\right)^{-9/4}\,,\\
    & = C + B\sum_{n=0}^8 \frac{c_n z^{n-5/4}}{n-5/4} F^{(3)}_D \left(n-\frac{5}{4};\frac{3}{4},\frac{9}{4},\frac{9}{4};n-\frac{1}{4};\frac{z}{Q} ,\frac{z}{\eta
    },-\frac{z}{\eta}\right)\,.\label{eq:gen_sol_m}
\end{align}


Since $E(r)$ is already known as an elementary analytical function of $r$, and $z(r)=r^2E(r)$, the mass function $m(r)$ can be expressed in closed form as a function of the areal radius $r$. The integration constant $C$ is fixed by imposing asymptotic flatness. As $r\rightarrow\infty$, one has $z\rightarrow Q$ and $m(r)\rightarrow M$, where $M$ is the ADM mass. Hence,
\begin{equation}
    C= M - B\sum_{n=0}^8 \frac{c_n Q^{n-5/4}}{n-5/4} F^{(3)}_D \left(n-\frac{5}{4};\frac{3}{4},\frac{9}{4},\frac{9}{4};n-\frac{1}{4};1 ,\frac{Q}{\eta
    },-\frac{Q}{\eta}\right)\,.\label{eq:const_C}
\end{equation}


\subsection{Exact Einstein--Euler--Heisenberg Case: $(b=7a/4)$}

In Einstein--Euler--Heisenberg (EEH) theory, the nonlinear couplings satisfy $b=7a/4$, and hence $c=2b-a=5a/2$. Substituting this relation into the quantities $C$, $B$, $c_n$, and $\eta$ appearing in Eq.~\eqref{eq:gen_sol_m}, we obtain the exact dyonic EEH black-hole solution
\begin{equation}
    m(r) = C + B\sum_{n=0}^8 \frac{c_n z^{n-5/4}}{n-5/4} F^{(3)}_D \left(n-\frac{5}{4};\frac{3}{4},\frac{9}{4},\frac{9}{4};n-\frac{1}{4};\frac{z}{Q} ,\frac{z}{\eta},-\frac{z}{\eta}\right)\,, \label{eq:EEH_exact_mass}
\end{equation}
where
\begin{equation}
    \eta^2= - \frac{5P^2}{2}\,, \qquad  B=\frac{1}{16\sqrt{2}(5a/2)^{9/4}P^{9/2}Q^{3/4}}\,.
\end{equation}
For $P>0$, the absolute value in the expression for $B$ may be omitted. The coefficients $c_n$ are obtained by substituting $c=5a/2$ into Table~\ref{tab:coeff}, while the integration constant \(C\) is determined by Eq.~\eqref{eq:const_C}. The auxiliary variable entering Eq.~\eqref{eq:EEH_exact_mass} is $z(r)=r^2E(r)$.

The purely electric limit $P\rightarrow0$ cannot be taken directly in Eq.~\eqref{eq:EEH_exact_mass}, because both $\eta$ and the denominator of $B$ vanish in this limit. We must instead return to the integral representation
\begin{align}
    \mathcal{I}(z) &= B\int dz \, \sum_{n=0}^8 c_n z^{n-9/4} \left(1-\frac{z}{Q}\right)^{-3/4} \left(1-\frac{z}{\eta}\right)^{-9/4} \left(1+\frac{z}{\eta}\right)^{-9/4}\,,\\
    &= B' \left(-\eta^2\right)^{-9/4}\int dz\, \sum_{n=0}^8 c_n z^{n-9/4} \left(1-\frac{z}{Q}\right)^{-3/4} \left(\frac{\eta^2-z^2}{\eta^2}\right)^{-9/4}  \,,\\
    &= B' \int dz\, \sum_{n=0}^8 c_n z^{n-9/4} \left(1-\frac{z}{Q}\right)^{-3/4} \left(z^2 - \eta^2\right)^{-9/4}  \,,\label{eq:reduced_I}
\end{align}
where the reduced prefactor is
\begin{equation}
    B' \equiv B \left(-\eta^2\right)^{9/4} = \frac{1}{16\sqrt{2} a^{9/4}Q^{3/4}}\,.
\end{equation}
Although $B$ is singular as $P\rightarrow0$, the combination $B(-\eta^2)^{9/4}$ remains finite. Eq.~\eqref{eq:reduced_I} therefore provides a regular representation of the purely electric limit.

When $P=0$, the integral no longer depends on $\eta$. Moreover, Table~\ref{tab:coeff} shows that only the coefficients with $n=6,7,8$ remain nonzero:
\begin{equation}
c_6=9a^2Q^2\,,
\qquad
c_7=-9a^2Q\,,
\qquad
c_8=2a^2\,.
\end{equation}
Consequently, Eq.~\eqref{eq:reduced_I} reduces to
\begin{align}
    \mathcal{I}_1(z) & = B' \int dz\, \sum_{n=6}^8 c_n z^{n-9/4} \left(1-\frac{z}{Q}\right)^{-3/4} \left(z^2\right)^{-9/4}  \,,\\
    & = B' a^2 \int dz\, z^{-3/4} \left(9Q^2-9Qz+2z^2\right) \left(1-\frac{z}{Q}\right)^{-3/4}\,. \label{eq:pure_electric_integral}
\end{align}
Thus, in the purely electric sector, the Lauricella function reduces to a sum of three Gauss hypergeometric functions.


To establish the connection with our previous purely electric solution, we adopt the parametrization introduced in Eq.~(78) of Ref.~\cite{Ahmed:2026ufj}, 
\begin{equation}
    \frac{z}{Q} = \frac{3w^{1/3}}{1+w^{1/3}+w^{2/3}}\,,\qquad
    1-\frac{z}{Q} = \frac{\left(1-w^{1/3}\right)^2}{1+w^{1/3}+w^{2/3}}\,,\qquad
    \frac{dz}{dw} = \frac{Qw^{-2/3}\left(1-w^{2/3}\right)}{\left(1+w^{1/3}+w^{2/3}\right)^2}\,,
\end{equation}
After substituting these relations into Eq.~\eqref{eq:pure_electric_integral}, the mass function becomes
\begin{equation}
    m(r) =C_{\mathrm{e}}
 + \frac{\left(27a Q^2\right)^{3/4}}{48\sqrt{2}a} \int dw \, (1-w)^{-5/2} w^{-1/4}(1+w) \left(w^{1/3}+w^{-1/3}\right) \left(w^{1/3}+w^{-1/3}-2\right)\,,
\end{equation}
where $C_{\mathrm{e}}$ is fixed by the asymptotic condition $m(r)\rightarrow M$ as $r\rightarrow\infty$. This expression coincides precisely with Eq.~(89) of Ref.~\cite{Luo:2026srx}, thus confirming that the exact dyonic solution consistently reproduces the previously obtained purely electric EEH black hole.


\subsection{The integrable locus: $(b=a/2)$}


The exact dyonic black-hole solution obtained in Ref.~\cite{Ahmed:2026ufj} corresponds to the special coupling locus
\begin{equation}
b=\frac{a}{2}\,,
\qquad
c=2b-a=0\,.
\end{equation}
As in the purely electric limit discussed above, the general Lauricella representation cannot be applied directly. Indeed, when $c=0$, one has
\begin{equation}
c_0=c_1=0,
\qquad
\eta=0,
\end{equation}
and the quantities $B$ and $\eta$ appearing separately in Eq.~\eqref{eq:gen_sol_m} become singular. Their combined limit, however, remains finite and is described by the reduced integral representation in Eq.~\eqref{eq:reduced_I}. Setting $c=0$, we obtain
\begin{align}
    \mathcal{I}_2(z)&= B' \int dz\, \sum_{n=2}^8 c_n z^{n-9/4}\left(1-\frac{z}{Q}\right)^{-3/4}\left(z^2\right)^{-9/4} \,,\\
    &= B' \int dz\, a^2 z^{-19/4}\left[-3P^4Q^2 + 5P^4Qz - 2P^4z^2 \right. \notag\\
    &\left.+ 6P^2Qz^3 + \left(9Q^2-4P^2\right)z^4 - 9Qz^5 + 2z^6\right]\left(1-\frac{z}{Q}\right)^{-3/4}\,. \label{eq:I2_z}
\end{align}


When $c=0$, Eq.~\eqref{eq:E_cubic1} reduced to
\begin{equation}
4aE(r)^3+E(r)-\frac{Q}{r^2}=0\,.
\end{equation}
Introducing the auxiliary variable $y$ through
\begin{equation}
y^2\equiv E(r),
\end{equation}
and recalling that $z=r^2E(r)$, the cubic equation gives
\begin{equation}
    \frac{z}{Q}=\frac{1}{1+4ay^4}\,,\qquad 1-\frac{z}{Q}= \frac{4ay^4}{1+4ay^4}\,,\qquad  \frac{dz}{dy}=-\frac{16aQy^3}{\left(1+4ay^4\right)^2}\,,
\end{equation}
Substituting these relations into Eq.~\eqref{eq:I2_z}, we find
\begin{equation}
     \mathcal{I}_2(z) = \int dy\,\left[-\frac{P^2\left(1+12ay^4\right)}{2\sqrt{Q}\sqrt{1+4ay^4}} + \frac{aP^4y^4 \left(1+12ay^4\right) (1+4ay^4)^{3/2}}{Q^{5/2}} - \frac{Q^{3/2}\left(1 +  6ay^4\right)\left(1 +  12ay^4\right)}{2\left(1+4ay^4\right)^{5/2}}\right]\,. \label{eq:I2_y}
\end{equation}
Following Eq.~(2.8) of Ref.~\cite{Ahmed:2026ufj}, the areal radius can be expressed parametrically as
\begin{equation}
    r^2=\frac{z}{y^2} = \frac{Q}{y^2\left(1+4ay^4\right)} \,,
\end{equation}
or equivalently,
\begin{equation}
    r=\frac{\sqrt{Q}}{y\sqrt{1+4ay^4}}\,, \end{equation}
Differentiating this relation yields
\begin{equation}
    \frac{dr}{dy}=-\frac{\sqrt{Q}\left(1+12ay^4\right)}{y^2(1+4ay^4)^{3/2}}\,.
\end{equation}
The first two terms in Eq.~\eqref{eq:I2_y} can then be recognized as total derivatives:
\begin{align}
    \frac{d}{dy}\left(\frac{P^2}{2r}\right) &= \frac{P^2\left(1+12ay^4\right)}{2\sqrt{Q}\sqrt{1+4ay^4}}\,,\\
    \frac{d}{dy}\left(\frac{aP^4}{5r^5}\right)&=\frac{aP^4y^4 \left(1+12ay^4\right) (1+4ay^4)^{3/2}}{Q^{5/2}}\,.
\end{align}
The mass function therefore takes the form
\begin{equation}
    m(r) = -\frac{P^2}{2r} + \frac{aP^4}{5r^5} -\frac{Q^{3/2}}{2} \int dy\, \frac{1+18ay^4+72a^2y^8}{\left(1+4ay^4\right)^{5/2}} \,. \label{eq:integrable_mass}
\end{equation}
The lower integration limit has been chosen so that $y\rightarrow0$ as $r\rightarrow\infty$, ensuring that $m(r)\rightarrow M$, with $M$ identified as the ADM mass. Eq.~\eqref{eq:integrable_mass} reproduces the exact result of Ref.~\cite{Ahmed:2026ufj}. Thus, the solution at the integrable locus $b=a/2$ is recovered as a regular limiting case of our general construction, even though the original Lauricella representation is singular when written directly in terms of $B$ and $\eta$.


\section*{Acknowledgment}
HL is supported by the Research and Practice Innovation Plan for Graduate Students in Jiangsu Province (No.~26CXJH7333). XYC is supported by National Science Foundation of China (No.~W2533026). DY was supported by the National Research Foundation of Korea (NRF) grant funded by the Korean government (No.~RS-2026-25476711). CC is supported by National Natural Science Foundation of China (No.~12503003, No.~1243300), National Key R\&D Program of China (No.~2021YFC2203100).


\begin{thebibliography}{200}

\bibitem{Luo:2026srx}
H.~Luo, N.~Cao, X.~Y.~Chew, K.~G.~Lim, C.~Chen and D.~h.~Yeom,
[arXiv:2607.21938 [gr-qc]].



\bibitem{Ahmed:2026ufj}
F.~Ahmed, A.~Al-Badawi and I.~Sakalli,
[arXiv:2608.23620 [gr-qc]].



\bibitem{Kraniotis:2010gx}
G.~V.~Kraniotis,
Class. Quant. Grav. \textbf{28}, 085021 (2011)
[arXiv:1009.5189 [gr-qc]].



\bibitem{Lai:2020mfi}
S.~H.~Lai, J.~C.~Lee and Y.~Yang,
Symmetry \textbf{13}, no.3, 454 (2021)
[arXiv:2012.14726 [hep-th]].



\bibitem{Duhr:2023bku}
C.~Duhr and F.~Porkert,
JHEP \textbf{02}, 179 (2024)
[arXiv:2309.12772 [hep-th]].



\bibitem{Akerblom:2004cg}
N.~Akerblom and M.~Flohr,
JHEP \textbf{02}, 057 (2005)
[arXiv:hep-th/0409253 [hep-th]].



\bibitem{Fathi:2023qyl}
M.~Fathi,
Annals Phys. \textbf{457}, 169401 (2023)
[arXiv:2305.09797 [gr-qc]].



\bibitem{Kaneko:2010gta}
T.~Kaneko,
PoS \textbf{CPP2010}, 010 (2010)
[arXiv:1105.2080 [hep-ph]].





\end{thebibliography}
\end{document}